\documentclass{aip-cp}

\usepackage[numbers]{natbib}
\usepackage{rotating}
\usepackage{graphicx}
\usepackage[utf8]{inputenc}

\newcommand{\ot}{\ensuremath{\frac{1}{2}}}

\newcommand{\be}{\begin{equation}}
\newcommand{\ee}{\end{equation}}
\newcommand{\bea}{\begin{eqnarray}}
\newcommand{\eea}{\end{eqnarray}}
\newcommand{\E}{\mathrm{e}}
\newcommand{\I}{\mathrm{i}}
\newcommand{\FD}{\;.}

\begin{document}

\title{Hadron Structure and Spectrum from the Lattice}

\author{C. B. Lang}
\eaddress{christian.lang@uni-graz.at}

\affil{Inst. of Physics, University of Graz, A-8010 Graz, Austria.}

\maketitle

\begin{abstract}
Lattice calculations for hadrons are now entering the domain of resonances and
scattering, necessitating a better understanding of the observed discrete energy
spectrum. This is a reviewing survey about recent lattice QCD results, 
with some emphasis on spectrum and scattering. 

\end{abstract}

\section{INTRODUCTION}

Quantum field theories in four dimensions are not well defined without some regularization. Wilson's \cite{Wilson:1974}
formulation on a Euclidean space-time lattice is such a regularization with the advantage of maintaining
gauge invariance and straightforward accessibility by computer. The path integrals become finite dimensional
integrals, however of very high dimensions. The continuum limit is obtained by keeping the physical volume fixed while letting
the lattice spacing $a$ approaching zero. The scale parameter is determined by comparing a physical observable
(e.g., a mass $m$) with the measured dimensionless lattice observable (e.g., the product $a\,m$). Once $a$ is determined, all further
lattice observables can then be translated to physical values. Since one has to fix also the quark mass parameters one has to trade $n_f+1$ physical quantities for the lattice scale and quark mass parameters.  In the continuum limit the lattice parameters are tuned
such that $a\to 0$ while keeping the physical quantities fixed. 

There are several concerns on the way to continuum results. The physical volume is limited to a few fermi, unless
the lattice is very coarse. Typical calculation have lattice spacings between 0.05 fm and 0.2 fm.
The leading finite size effect is due to the lightest hadron, the pion, thus one wants  lattice
sizes $L$ where $L m_\pi$ is large; typical values are larger than 4. Below that value boundary effects may be sizeable.
This gives $L>6$ fm and leads to lattices of demanding $100^4$ sites.
Bringing the quark masses and, equivalently the pion mass, down to physical values necessitates large volumes or good
control on the finite size dependence and of the scaling behaviour in $a$. 

The main  object of lattice simulations are correlation functions, the Euclidean equivalent of n-point functions.
Masses or better: energies are obtained from the exponential decay of hadron propagators. However, in a quantum channels there
will be contribution of (formally: infinitely) many states:
$C_{ij}(t)\equiv \langle X_i(t)\overline X_j(0)\rangle=
\sum_n \langle X_i |n\rangle\, \E^{-t \,E_n}\,\langle n| \overline X_j\rangle$.
Due to the finiteness of the lattice volume the energies are discrete even in the situation of
open scattering channels. Asymptotically, for large $t$ the ground state dominates. However, the statistical
errors increasingly obscure the signal with increasing $t$ and most often one
has to work at not so large $t$.  

Depending on the type of calculation the excited states may be a nuisance or 
an advantage. In case one is interested in hadronic ground state parameters (there are not many
such hadrons: the pseudoscalars and the nucleon with its strange, charmed, beautiful and maybe topped cousins)
or in 3-point functions (like form factors or other matrix elements) the  excited states are a ``contamination'' and
one fights to get rid of their influence. If, on the other hand, one is interested in excited hadrons and decay properties one needs
the excitation energies as precise as possible.

\section{HADRON STRUCTURE}

Obviously I have to concentrate here on a few highlights and the choice is subjective. Recent more complete reviews on the topic are \cite{Constantinou:2014tga,Zanotti:2015}. Also I am considering only information available at the time of this conference (Sept. 2015) with emphasis on results in the recent two years.

\begin{figure}
\centerline{
\includegraphics[width=4.5cm,angle=90]{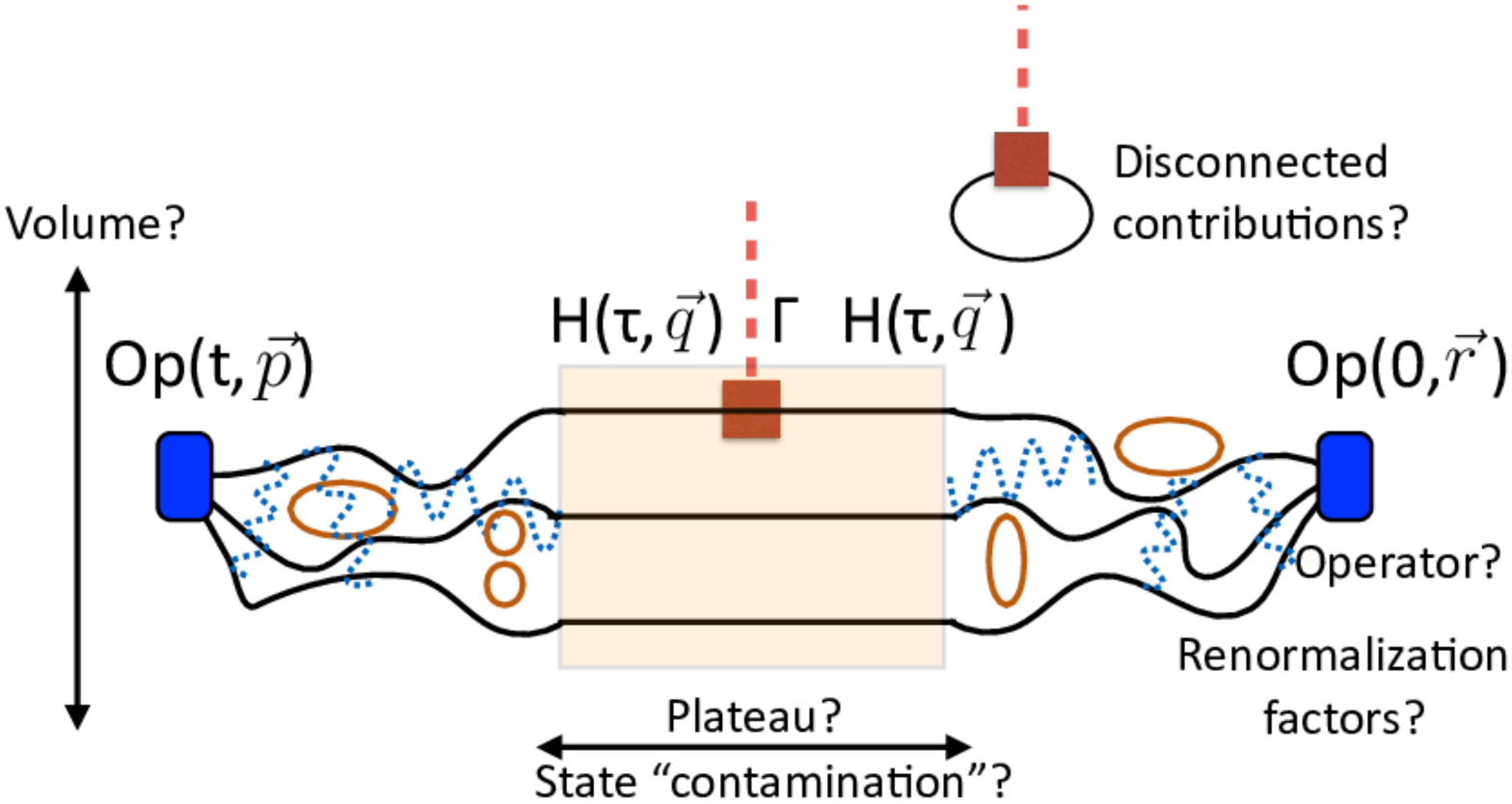}
\hfil
\includegraphics[width=7.5cm]{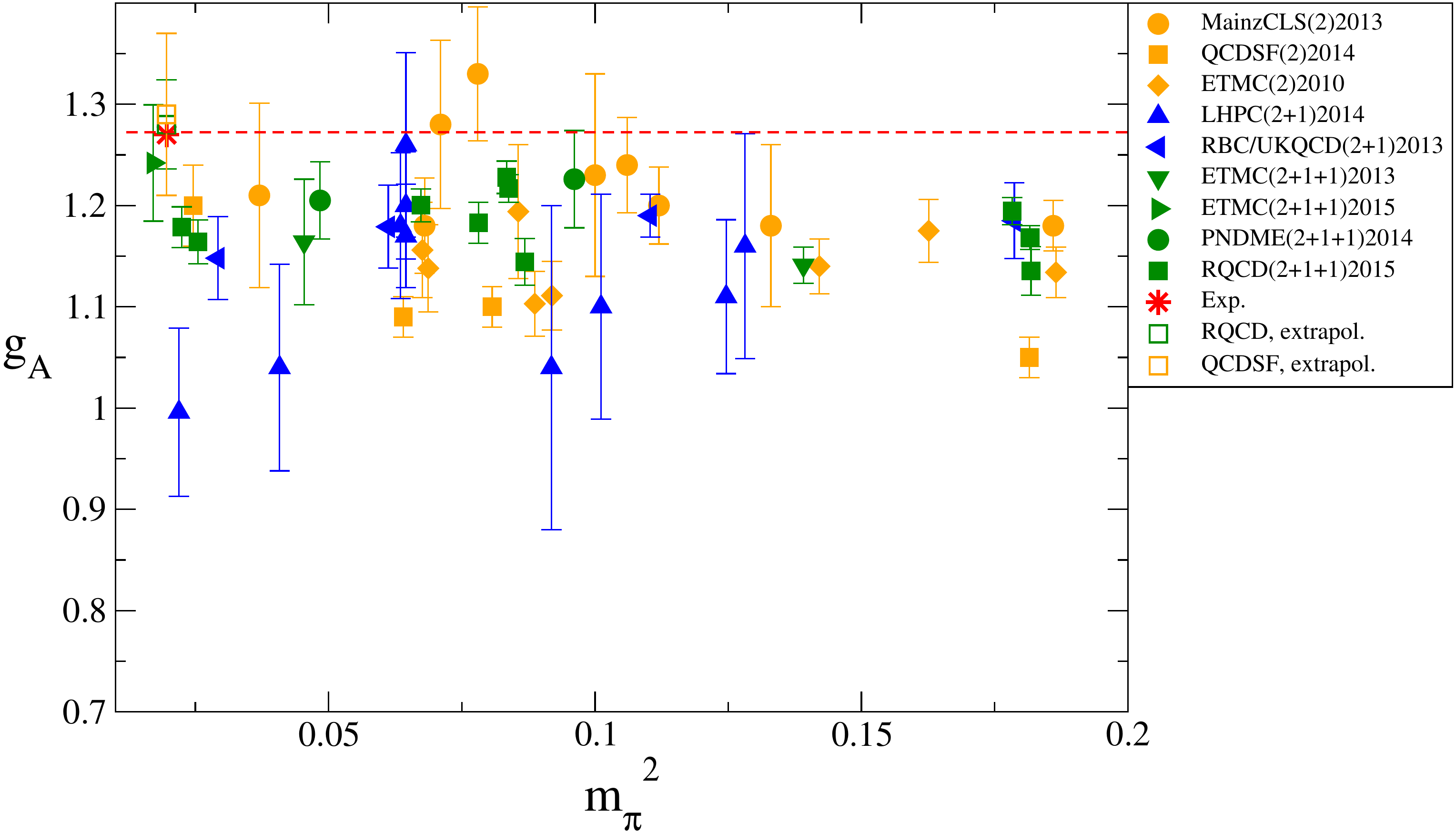}}
\caption{
\label{fig:matrixelement}Left: Matrix element calculation.
\label{fig:gA}Right:  Axial charge of the nucleon, comparison of values obtained by several groups.
The color indicates groups using the same number
of dynamical fermions ($N_{u,d}+N_s+N_c$); no significant differences are seen.}
\end{figure}

Hadron structure calculations are based on studying 3-point functions. A current is inserted between
an incoming and an outgoing hadron
$\langle H|\Gamma|H'\rangle$. In terms of lattice operators $O$ (also called interpolators) we have
\be
\langle O(t,\vec p)|\Gamma(\tau)|\overline O(0,\vec r)\rangle\FD
\ee
Fig. \ref{fig:matrixelement} shows the situation (and the concerns).
Assuming that the temporal distances $|\tau|$ and  $|t-\tau| $ are large enough that the
ground state hadron dominates the intermediate state,
\be
G_3\equiv\langle O(t,\vec p)|H(\vec p)\rangle \frac{e^{-E_p (t-\tau)}}{2E_p}
\mathbf{\langle H(\vec p)|\Gamma(\tau)|H(\vec r)\rangle}
\frac{e^{-E_r \tau}}{2E_r}\langle H(\vec r)|\overline O(0,\vec r)\rangle
\ee
the components factorize. One also determines the 2-point function
\be
G_2=\langle O(t,\vec q)\overline O(0,\vec p)\rangle
=\langle O(t,\vec p)|H(\vec p)\rangle \frac{e^{-E_p t}}{2E_p}\langle H(\vec p)|\overline O(0,\vec r)\rangle
\FD
\ee
and retrieves  the wanted matrix element $\langle H|\Gamma|H\rangle$ from a plateau behavior of suitable ratios
of $G_3$ and $G_2$.

In lattice calculations of matrix elements there are several concerns:
volume (finite size effects), lattice hadron operators, contamination from excited states, 
disconnected contributions (depending on the insertion type), renormalization factors (and possible mixing with other operators), dependence on
the pion (quark) mass and lattice spacing. Most studies are at higher pion mass and have to be extrapolated to the physical value. All these aspects have to be carefully considered for a reliable
result.

\subsubsection{Axial Charge of the Nucleon}

That axial isovector coupling $g_A$ can be obtained from $\langle p|\bar u \gamma_\mu \gamma_5 d|n\rangle$.
Fig. \ref{fig:gA}, right\footnote{Thanks to Martha Constantinou and Sara Collins for help.} show results of recent years. All are slightly below the experimental value $g_A/g_V=1.2723(23)$. The different numbers of dynamical quarks do not explain this. The dependence on lattice spacing or volume (recent results cover a range $3<m_\pi L<6$)
also shows no trend (cf. plots in \cite{Constantinou:2014tga}).

The most likely suspect is the influence of excited states that may be still significant in the
region of the insertion. Fitting the mentioned ratio at several values of $\tau$ to a plateau may lead to an
underestimation of the value. Analysis variants are a summation method \cite{Capitani:2010sg} or adding
excited states to the fit function. Recent studies carefully analyse those approaches \cite{Abdel-Rehim:2015owa,Bali:2014nma}.
From Fig. 2 of Ref. \cite{Bali:2014nma} one clearly sees that the nucleon has admixture from excitations up to about
0.6 fm; the source and sink are at $t=0$ and $t=t_f\approx 1$ fm, so the center of the hoped-for plateau
is at 0.5 fm, clearly in the contaminated region. In the ratio $g_A/F_\pi$ ($F_\pi$ is the pion decay constant)
some of the finite volume influence seems to cancel \cite{Horsley:2013ayv} and extrapolation to physical pion masses gives a value
close to experiment \cite{Bali:2014nma}.

Further recent results include a study of the disconnected contribution to the isoscalar (S and A) matrix elements, which are O(7\%)
\cite{Abdel-Rehim:2013wlz}. In  \cite{Bali:2014nma,Abdel-Rehim:2015owa}
also isovector couplings $g_S$ and $g_T$ have been determined and 
in a ChPT study the nucleon-pion-state contributions in the determination of the nucleon axial charge have been estimated to be a few percent \cite{Bar:2015zha}.

\subsubsection{Nucleon Electromagnetic Form Factors}

The so-called Dirac ($F_1$) and Pauli ($F_2$) form factors are determined from the matrix element
\be \langle N(p)|V_\mu |\overline N(r)\rangle\sim
 \bar u_N(p)\left[F_1(q^2)\gamma_\mu +F_2(q^2)\frac{\I \sigma^{\mu\nu} q_\nu}{2m_N}\right] u_N(r)
 \quad\textrm{with}\quad q^2\equiv(p-r)^2\equiv -Q^2\;.
 \ee
Here recent work has been already at close to physical pion masses \cite{Green:2014xba,Capitani:2015sba,Abdel-Rehim:2015jna,Yamazaki:2015vjn}. The results are generally consistent. They cover, however, only a very small range of $Q^2$ as compared to experiments. The reason is indigenous to the lattice approach. Due to the finite box size the momenta are quantized. (E.g., $q^2=\vec k^2 (2\pi/L)^2$ for the non-interacting case, where $\vec k$ is a vector with integer components.) This constrains both, the lowest and the highest achievable values. For small $q^2$ one need large volumes, for large $q^2$ the
statistical noise increases. Already $\vec k^2=6$, which corresponds to $Q^2\approx 1 $ GeV$^2$ for $L=3$ fm, is a problem in that regard. A  recently proposed approach utilizing the Feynman-Hellmann 
relation between $\langle h|\mathcal{O}|h\rangle$ and the derivative of a 2-point function may help in going to larger $Q^2$ \cite{Chambers:2014qaa,Chambers:2015kuw,Shanahan:2014tja}

\begin{figure}

\begin{minipage}{0.5\textwidth}
\includegraphics[width=6cm,clip]{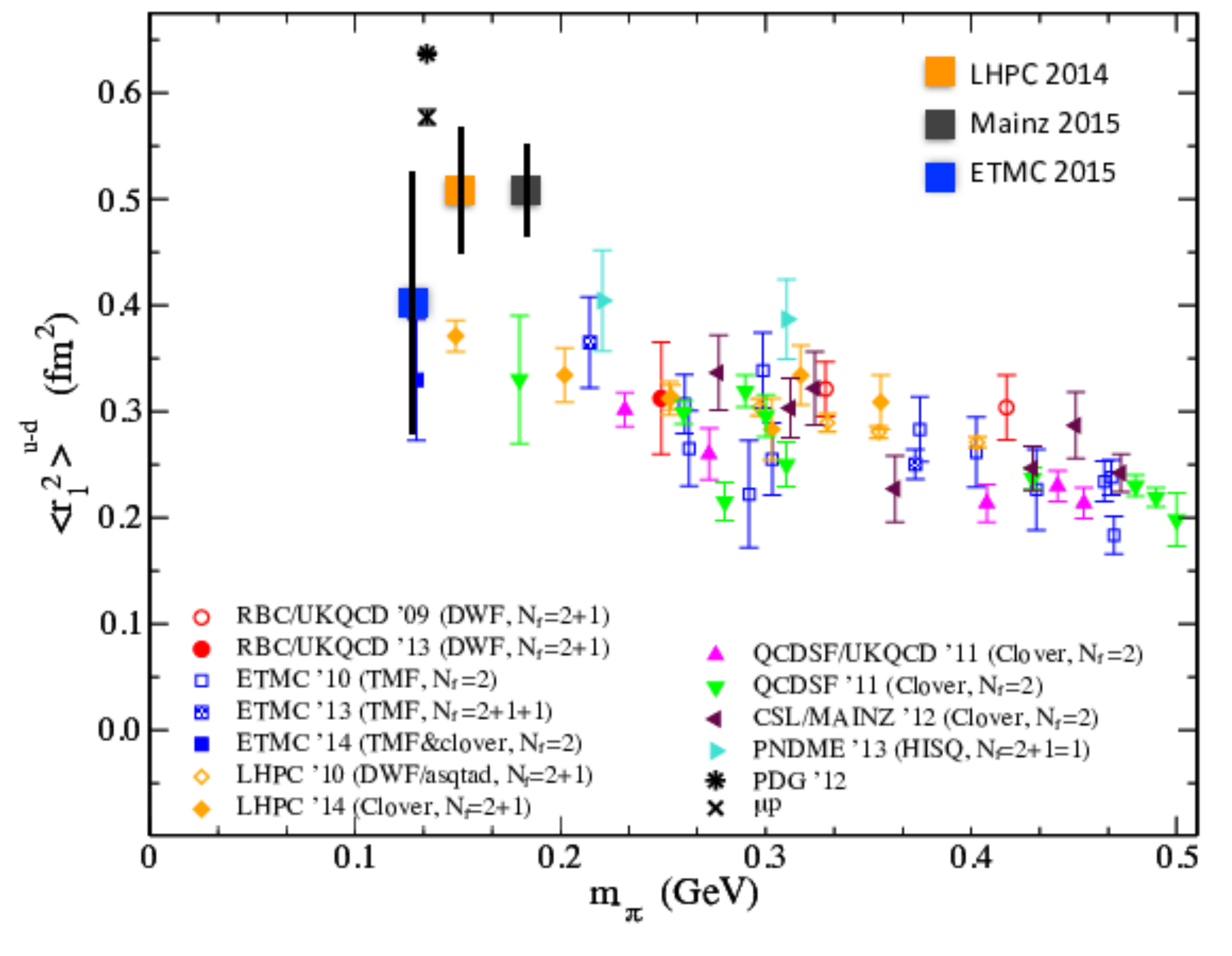}
 \end{minipage}
\begin{minipage}{0.4\textwidth}
\begin{tabular}{lll}
\hline\vspace*{-12pt}\\\small
	&$\mathbf{m_\pi}$\bf{ [MeV]}&$\mathbf{(r_1^2)^v}$\bf{[fm}$\mathbf{^2}$\bf{]}\\
\hline\vspace*{-10pt}\\
LHPC \cite{Green:2014xba}&149&	0.498(55)  \\
Mainz CLS \cite{Capitani:2015sba}&193&	0.501(42) 	\\
ETMC \cite{Abdel-Rehim:2015jna}&135&    0.398(126)\\
\hline
Exp. $ep$ \cite{Mohr:2012tt}&      &0.640(9)     \\
 Exp. $\mu p$ \cite{Antognini:1900ns}       &      & 0.578(2)    \\
\hline
\end{tabular}
\end{minipage}

\caption{\label{fig:chargeradius}Dirac charge radius $r_1$ from lattice data (Figure from \cite{Constantinou:2014tga} with superimposed new points due to \cite{Green:2014xba,Capitani:2015sba,Abdel-Rehim:2015jna}). The Table compares the results with the numbers from $e\,p$ scattering and muonic hydrogen experiments (the difference in the numbers from the experiments is not yet understood).}
\end{figure}

Lower values of $Q^2$ are important for the charge radii which are obtained from extrapolating fits to the lattice data. As can be seen from Fig. \ref{fig:chargeradius}  the bulk of the
values is significantly  smaller than both experimental values.

\subsubsection{The Proton Spin}

The proton spin has contributions from the quarks and the gluons: $\frac{1}{2}=\sum_q J_q\;+\;J_G$.
where one may distinguish the quark orbital and spinor contributions $J_q=L_q+\ot\Delta\Sigma^q$,
suggested in \cite{Ji:1996ek}. The quark contributions need computation of 
matrix elements like $\langle x \rangle^q$ and $\langle p'|T^{\mu\nu}|p\rangle$, involving  derivative operators.
To determine individual $\Delta\Sigma^q$ one needs to consider disconnected contributions which require high statistics and special methods (stochastic source methods).

All lattice calculations need to be extrapolated down to the physical pion mass. In \cite{Alexandrou:2013joa} Heavy Baryon Chiral Perturbation Theory was used and the results are
quite sensitive on the extrapolation leading to large systematic errors. For the light quarks the values have stabilized at $\Delta u+\Delta d=0.35(6)$ 
\cite{Alexandrou:2013joa,Yang:2015xga,Constantinou:2014tga}.
For the strange quarks the contribution comes from gluonic coupling to vacuum loops. Several
collaborations' results \cite{Zanotti:2015} are barely dependent on $m_\pi$  giving a value
$\Delta s=-0.02(1)$ \cite{Chambers:2015bka}. This sums to 
$\Delta \Sigma=\Delta u+\Delta d+\Delta s=0.33(7)$ in good agreement with the 
COMPASS(2007) value 0.33(3)(5).

The uncertainties are still large: the pion mass extrapolation introduces significant model dependence. For a recent more detailed review see \cite{Liu:2015nva}.

\subsubsection{Low Energy Parameters}

Low energy parameters like leptonic and semileptonic decay constants, CKM matrix elements, quark masses, the quark condensate, $\alpha_s$ and others are collected in the
compilation of the Flavor Lattice Averaging Group - FLAG (\verb+http://itpwiki.unibe.ch/flag+) 
\cite{Aoki:2013ldr}. Heavy meson decay constants can be found in recent work by \cite{Colquhoun:2015oha}.

\subsubsection{Radiative Decays}

On the lattice on-shell decays are forbidden due to the Maiani-Testa theorem; however
Lellouch and L\"uscher \cite{Lellouch:2000pv}
found a method to circumvent that problem. Recently Brice\~no \cite{Briceno:2014uqa} formulated a technique to address readiative decays like $\rho\to\pi\gamma^*$ (For alternative approaches see \cite{Bernard:2012bi,Agadjanov:2014kha}). In \cite{Shultz:2015pfa} the $\rho$ was assumed to be stable and
basic tools for the analysis were formulated. The pion mass there is quite high of $\mathcal{O}(700$ MeV). The CSSM collaboration presented results at almost physical pion mass (157 MeV)\cite{Owen:2015fra}.

In the real process, however, the $\rho$ is a resonance: $\pi\gamma^*\to\rho\to\pi\pi$. This now has been studied in a lattice simulation \cite{Briceno:2015dca}. The transition matrix element was computed and a parametrisation of the amplitude allows the analytic continuation
to the $\rho$-pole in the unphysical sheet and extraction of the form factor $F_{\pi\rho}(E_{\pi\pi}^*,Q^2)$ from the residue. The calculation still is for large pion mass of 400 MeV but compares favorable with phenomenological model calculations. For more information see Brice\~no's contribution to this conference.

\section{HADRON SPECTROSCOPY}

\subsection{Single Hadron Approximation}

A recent highlight is the determination of the electromagnetic mass differences for
$p$, $\Sigma$, $\Xi$ and others \cite{Borsanyi:2014jba}. Four quark species $u$, $d$, $s$, and $c$
were taken into account and QED in its non-compact version was added to QCD, both
non-perturbatively. QED needs special care: gauge fixing, finite volume corrections 
$\mathcal{O}(1/L)$ and regularisation scheme make life hard (see also \cite{Davoudi:2014qua,Endres:2015gda}). The results obtained for 197 MeV $\leq m_\pi\le$ 440 MeV were extrapolated
to the physical point leading to high precision values in good agreement with experiment, in some cases predictions like for $\Delta\Sigma$.

Milestones in the determination of the hadronic states were \cite{Durr:2008zz,Edwards:2011jj,Edwards:2012fx}. In \cite{Durr:2008zz} prominent members of the ($u$, $d$, $d$) family of hadrons were obtained, in \cite{Edwards:2011jj,Edwards:2012fx} meson and baryon excitations were determined for several spin-parity channels. 
This year has brought results on singly- and doubly charmed baryons with and without strangeness\cite{Bali:2015lka} for ground states and first excitations. The pion masses were between 260 and 460 MeV and the results could be extrapolated to the physical point. 
Ground state energies for baryons with up to three heavy quarks ($c$ and $b$)  were computed by \cite{Brown:2014ena}  at several pion masses and extrapolated to the physical point.

A challenging problem is the identification of spin, since different continuum spins
couple to the same lattice operator. Comparing the overlap patterns the Hadron Spectrum Collaboration resolved spins up to 4 in the (excited) charmonium spectrum \cite{Liu:2012ze} and for charmed mesons \cite{Moir:2013ub}. Based on that experience, in \cite{Padmanath:2015jea} doubly charmed baryons were studied (for a pion mass of 400 MeV) and spin identification up to 7/2 was performed. The baryon lattice operators were constructed by subduction of continuum operators to the lattice symmetry. Up to 11 excitation energy levels were presented; the observed multiplet structure matches the non-relativistic quark spinor model with symmetry $SU(6)\times O(3)$.
 
The bulk of results was extracted from correlation functions of single hadrons, i.e., either baryonic three-quark operators or mesonic quark-antiquark operators. Although we know that in quantum field theory all possible multi-quark intermediate states can contribute due to the dynamical vacuum with fermion loops, in practical lattice calculations using single hadron correlations these contributions are suppressed. This explains why one finds signals for resonances although they are not asymptotic states. The influence of coupled channels and associated thresholds is effectively neglected. It also means that an observed excited level do not necessarily give the position of the resonance peak. One has to allow for multi-hadron operators in the set of lattice interpolators.

\subsection{Multi-Hadron Approach}

This led to a changed point of view: One does not study the resonance correlators but the scattering process where resonances may appear. Due to the finite volume the energy spectrum of
the scattering process is discrete. L\"uscher derived a relation \cite{Luscher:1986pf,Luscher:1990ux} between the spectrum
in finite volume to the phase shift in the continuum for elastic meson-meson scattering. This has
been extended to moving frames and hadron-hadron scattering in general. In recent years there
has been an explosion of contributions in that direction.

What are the challenges in that approach? First one needs to consider a larger set of operators -
single hadron as well as hadron-hadron operators - and cross-correlations $C_{ij}(t)$ between them.
In that correlation elements one has (for baryon-baryon scattering) up to six valence quark propagators. Secondly, there
will be quark-antiquark annihilations (``backtracking quarks'') in disconnected or partially
disconnected terms. This is a notorious problem in such simulations and needs high statistics and
efficient new tools like stochastic sources or distillation. In the distillation method
\cite{Peardon:2009gh,Morningstar:2011ka} the hadron operators are constructed from quark sources that are eigenvectors of the spatial Laplacian. Once the quark propagators between these sources, the so-called perambulators, have been constructed, it is possible to efficiently compute correlations between
different operators. Changing the operators and projection to momenta can be done independent of the perambulators and so the method is very versatile.

L\"uscher's original method was valid in the elastic region but meanwhile there are extensions to several coupled channels \cite{Bernard:2010fp} including nucleon-nucleon scattering, moving frames and arbitrary spin \cite{Briceno:2012yi,Briceno:2013lba,Briceno:2014oea} and generalizations of the Lellouch-L\"uscher $1\to 2$ transitions \cite{Hansen:2012tf,Briceno:2014uqa}.

Often the decay is (like, e.g., in $a_1\to \rho\pi\to\pi\pi\pi$) to a three-hadron state and also there theoretical results were presented recently \cite{Hansen:2014eka,Meissner:2014dea,Hansen:2015zga}. No actual lattice simulation exists yet.

Following several studies of elastic $\pi \pi$ and $\pi K$ scattering as well as coupled channels
model calculations \cite{Bernard:2008ax,Guo:2012hv} the last year has finally brought the first
coupled channel simulation. Dudek et al. \cite{Dudek:2014qha} investigated $s$-, $p$- and $d$-waves
of the coupled $\pi K -\eta K$ system. Three lattice sizes with up to 70 identified energy levels and an interpolating model allows the determination of phase shifts and inelasticity up to 1600 MeV. Due to the large pion mass of 391 MeV the $K^*$ comes as
a bound state but in particular in $s$ and $d$ wave the main features are successfully reproduced.
This promising result was followed by a $\pi\pi$, $K\overline{K}$ coupled channel study \cite{Wilson:2015dqa}. These results for a pion mass of 236 MeV have then been extrapolated to the physical point \cite{Bolton:2015psa}. More details can be found in the contributions of J. Dudek, D. Wilson and 
D. Bolton to this conference.

A first application of $\pi N$ scattering was presented already two years ago \cite{Lang:2012db, Verduci:2014csa} demonstrating the importance of scattering states. Earlier results for the
$\ot^-$ channel did show two energy levels tentatively attributed to the $N(1535)$ and $N(1650)$, but
the splitting was too large. In the new study the lowest $\pi N$ $s$-wave level was correctly  identified closely below
threshold  and the next two level had the right splitting and position of $N(1535)$ and $N(1650)$.
Meanwhile further results with multi-hadron interpolators have appeared \cite{Kiratidis:2015vpa}.

Nucleon-nucleon scattering needs six valence quark propagators but none of the quarks  is backtracking.
Such a study needs large spatial lattice size; results for $s$, $p$, $d$, and $f$ partial waves and spatial
extent 4.6 fm was presented recently \cite{Berkowitz:2015eaa}. The pion mass there
is quite high (800 MeV) but further studies closer to the physical values are to be expected.

\subsection{Heavy Quarks}

Recent reviews on lattice results in the heavy flavor sector are
\cite{Prelovsek:2015fra,Mohler:2015zsa,Padmanath:2015bra}.
At present it is hopeless to perform a full coupled channel phase shift lattice calculation in the charmonium sector - there are too many coupled channels in the interesting energy regions. 
On the way towards that far-lying goal we can, however, learn something from the
measured energy levels. An example for this ``level hunting'' is the search for a signal of the 
$Z_c^+(3900)$ state. In \cite{Prelovsek:2014swa} 18 interpolators of meson-meson type with $c\overline{c}u\overline{d}$ quark content as well as four tetraquark operators were included in the cross-correlation matrix. All observed levels (covering the energy range up to 4.1 GeV) could be identifed with (expected) meson-meson states and no extra state (which then could be associated to the $Z_c^+(3900)$) was found in this $I^G(J^{PC})=1^+(1^{+-})$ channel. This agrees with other lattice studies \cite{Lee:2014uta,Chen:2014afa}. There is an ongoing discussion whether the $Z_c(3900)$
might be a threshold effect. This has been also discussed in
the so-called HALQCD approach \cite{Ikeda:2015xxx}. There a potential related to the Nambu-Bethe-Salpeter equation is determined in a coupled channel formalism \cite{Aoki:2012bb}.

Charmonium levels in the single hadron approximation are in good agreement with experiment only below the $D\overline{D}$ threshold. In \cite{Lang:2015sba} charmonium $\psi(3770)$ was studied in a system of 15 operators of $c\overline{c}$ type as well as two $D\overline{D}$ interpolators for two pion masses (266 MeV and 157 MeV). Below threshold $\psi(2S)$ and above threshold $\psi(3770)$ were identified, both in good agreement with experiment.

\begin{figure}
\centerline{
\hfil\includegraphics[height=2.5cm,clip]{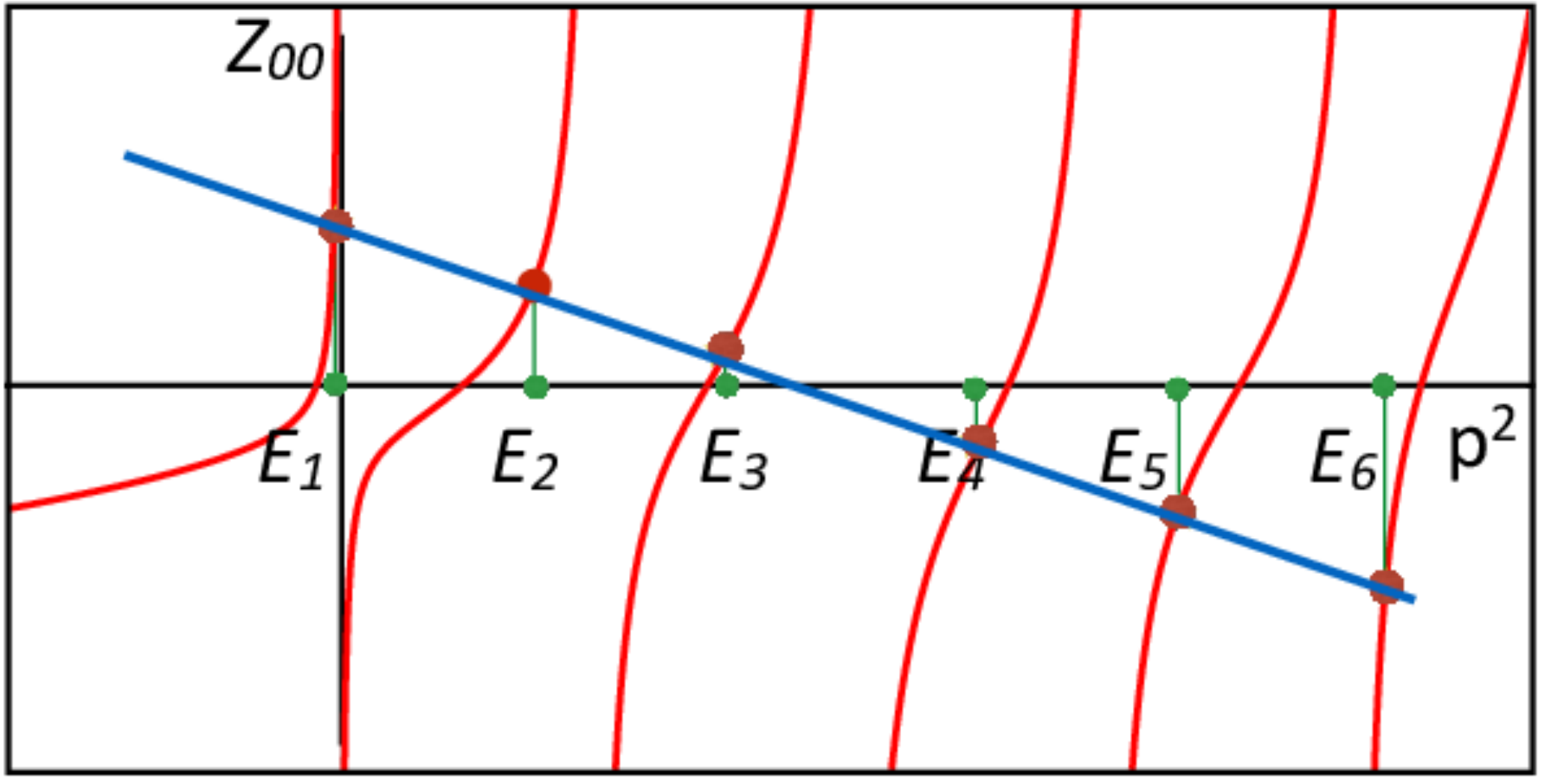}\hfil
\includegraphics[height=2.5cm,clip]{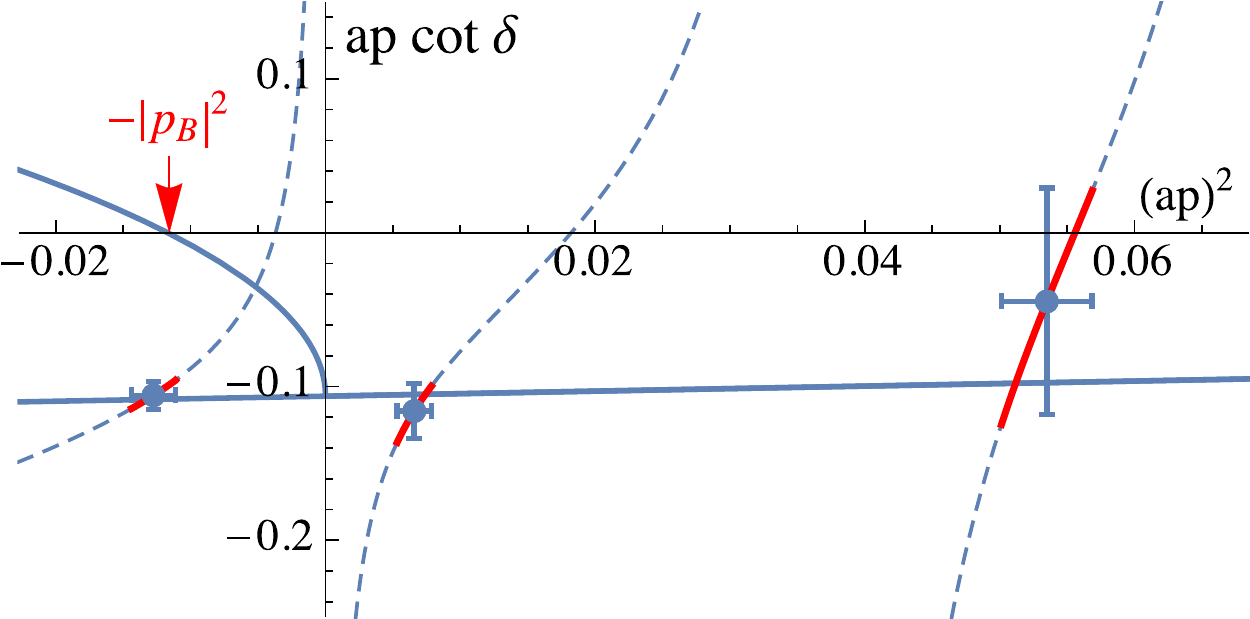}\hfil}
\caption{
\label{fig:Luscheranalysis}Left: Schematic description of the L{\"u}scher analysis: Red curves are the theoretically possible values, the measured energy values then lead to the values of $k^{-1}(s)$ lying on that curves. Right: Example for this scenario for $B_s$ ($0^+$) $BK$ scattering (Fig. from \cite{Lang:2015hza}); note that below threshold the analytic continuation of the phase space factor contributes to the real part leading to the bound state position.}
\end{figure}

Of particular interest are resonances or bound states close to thresholds.  The reciprocal partial wave scattering amplitude (in the elastic regime) may be parametrized by
\be
\mathrm{Re}[f_\ell^{-1}(s)]=\rho(s)\cot \delta_\ell(s)-\I\rho(s)\equiv k^{-1}(s)-\I\rho(s)
\quad\textrm{with the phase space factor}\quad \rho(s)=2p^{2\ell+1}/\sqrt{s}
\ee
and in the L\"uscher-type analysis each energy level gives a value of 
$\mathrm{Re}(f^{-1})=\,c \,\mathcal{Z}_{00}\left(1;\left(\frac{pL}{2\pi}\right)^2\right)$ (cf. Fig. \ref{fig:Luscheranalysis}). Above threshold each point gives a value of the phase shift. Interpolation and continuation below threshold allows to retrieve  threshold parameters as well bound state energies or resonance position and coupling.

\begin{figure}
\centerline{
\hfil\includegraphics[height=5cm,clip]{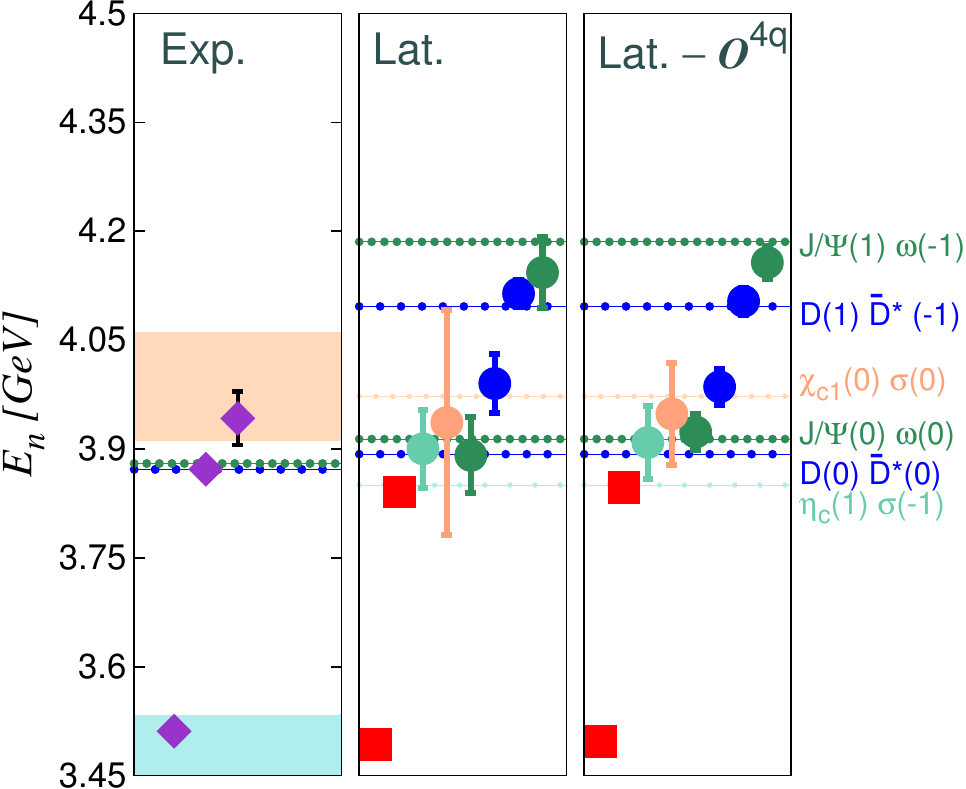}
\hfil\includegraphics[width=5cm,angle=90,clip]{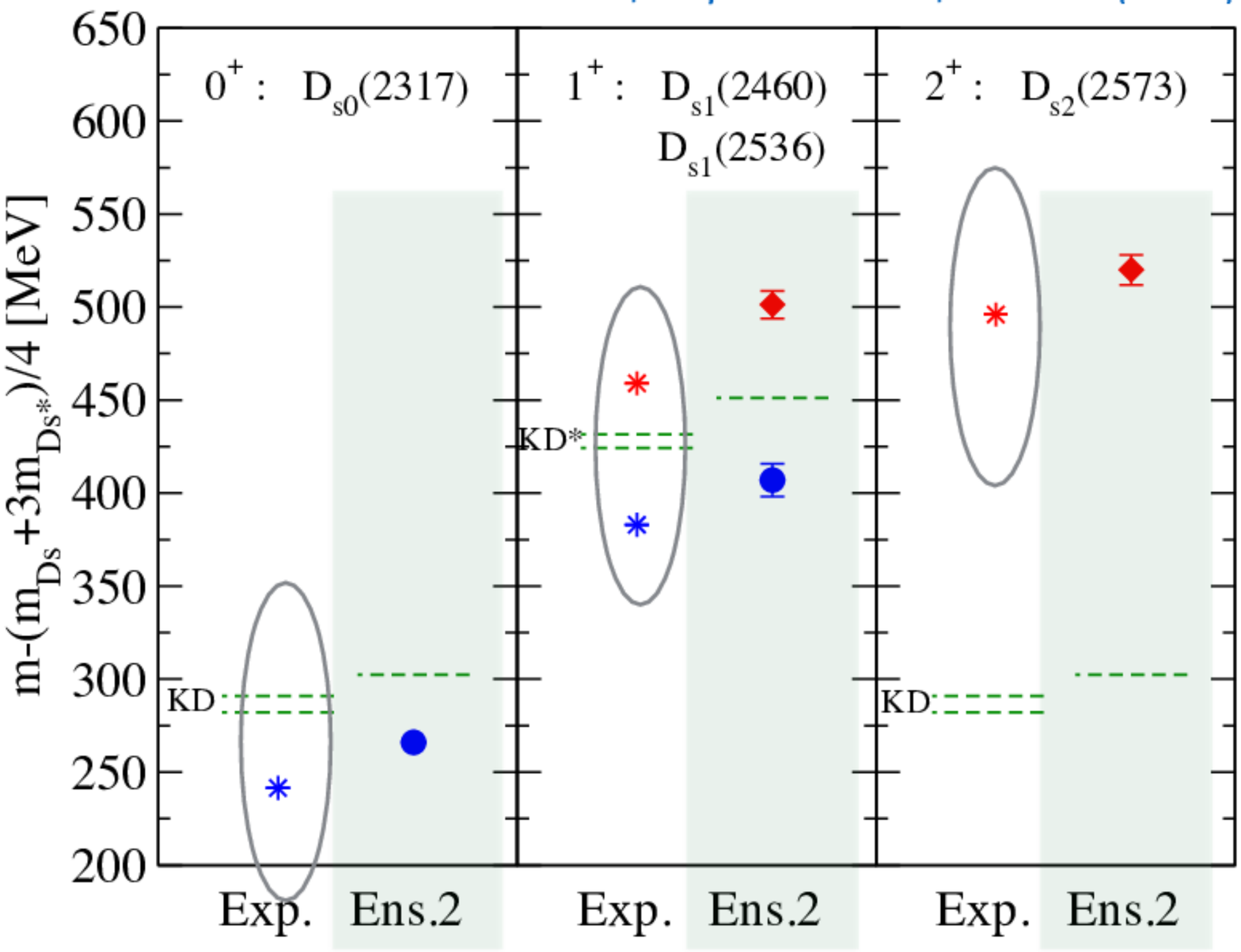}\hfil
}
\caption{
\label{fig:X3872}Left: Fig. from \cite{Padmanath:2015era} for the $I=0$   channel with $c\overline{c}$ and $c\overline{c}(u\overline{u}+d\overline{d})$ interpolators. The tetraquark operators appear to have little effect on the spectrum. The red squares are dominated by $c\overline{c}$ operators and are attributed to $\chi_{c1}$ and $X(3872)$. Right: Fig. from \cite{Lang:2014yfa}; results for $D_s$ states derived for a gauge field ensemble ``Ens.2'' with pion mass of 157 NeV \cite{Aoki:2008sm}.}
\end{figure}

In the $0^+(1^{++})$ channel lies the $X(3872)$; this state was postdicted in a lattice study (for the first time) \cite{Prelovsek:2013cra}. This was confirmed in \cite{Lee:2014uta}. A  recent study
\cite{Padmanath:2015era} extended the set of coupled channel operators significantly (22 interpolators including $DD^*$,
$J/\psi \omega$,  $\eta_c \sigma$, $\chi_{c1}\sigma$, as well
as four tetraquark operators). The $X(3872)$ closely below $DD^*$ threshold was reconfirmed with a
strong $c\overline{c}$ Fock component. It is not seen, if the $c\overline{c}$ interpolators are not included.

Phenomenological models as well as lattice calculations gave  controversial results for the
$D_s$ in $0^{++}$ and $1^{++}$. In both cases there is a nearby threshold: $KD$ and $KD^*$, respectively, and it was suggested that these channels may be important components of the states
\cite{vanBeveren:2003kd}. Indeed a lattice study \cite{Mohler:2013rwa,Lang:2014yfa} including these channels reproduced the
pattern from experiment and identified bound states $D_{s0}(2315)$ and $D_{s1}(2460)$ and, above the $KD^*$ threshold $D_{s1}(2536)$ (Fig. \ref {fig:X3872}). The levels were consistently higher than experiment due to
larger than physical pion mass of 157 MeV and quark mass tuning effects but the splitting
and distance to threshold agreed with experiment.

Motivated by these results a similar study was then done for $B_s$ in $0^{+}$, $1^{+}$ and $2^{+}$ with $BK$ and $B^*K$ contributions \cite{Lang:2015hza}. In $0^+$ a bound state $B_{s0}$ with a mass of 5.711(13)(19) GeV and
in $1+$ a bound state $B_{s1}$ with a mass of 5.750(17)(19) GeV was predicted. Close to
threshold a  weakly coupled $B_{s1}^0$ at a mass of 5.831(9)(6) GeV was identified close to
the experimental state at 5.8288(4) GeV.

\subsection{Summary}
 
The lattice formulation of QCD is mathematically well defined and provides a controlled
continuum limit. With increasing compute power and algorithmic improvements we have come close to that ambition. Lattice structure results approach the quality needed for an input to experiment analysis, although they are not yet precise enough and one still has to understand the origin of deviations. Efficient methods for disconnected graph contributions are needed. Our understanding of lattice scattering has improved considerably and hadron spectroscopy has entered a new era. Processes involving several coupled channels are still a challenge.

\section{ACKNOWLEDGMENTS}

Many thanks go to my collaborators of recent years Sasa Prelovsek, Daniel Mohler,  Luka Leskovec and Padmanath Madanagopalan. 
I thank Sara Collins  and Martha Constantinou for their help with the data collection and
Constantia Alexandrou, Raul Brice\~no, Christine
Davies, Michael Engelhardt, Gian-Carlo Rossi and Andr\'e Walker-Loud for information.
Support by the Austrian Science Fund FWF: I1313-N27  is gratefully acknowledged,



\end{document}